\documentclass[iop]{emulateapj}

\usepackage{graphicx}
\usepackage{apjfonts}
\usepackage{natbib}
\usepackage[colorlinks,citecolor=blue,linkcolor=blue,urlcolor=blue,dvips]{hyperref}

\newcommand{\phcms}{${\rm photons}\;{\rm cm}^{-2}\;{\rm s}^{-1}$} 
\newcommand{\Fermi}{{\it Fermi}}

\newcommand{\Swift}{{\it Swift}}

\newcommand{\mev}{MeV}

\slugcomment{Accepted for Publication in ApJL}

\shorttitle{Minute-timescale $\gamma$-ray variability of quasar 3C~279 in 2015 June}
\shortauthors{\Fermi-LAT Collaboration}

\begin{document}


\pagenumbering{arabic}

\title{Minute-Timescale $>100$\,\mev\ $\gamma$-ray variability during the giant outburst of quasar 3C~279 observed by \Fermi-LAT in 2015 June}


\author{
M.~Ackermann\altaffilmark{1}, 
R.~Anantua\altaffilmark{2}, 
K.~Asano\altaffilmark{3}, 
L.~Baldini\altaffilmark{4,2}, 
G.~Barbiellini\altaffilmark{5,6}, 
D.~Bastieri\altaffilmark{7,8}, 
J.~Becerra~Gonzalez\altaffilmark{9,10}, 
R.~Bellazzini\altaffilmark{11}, 
E.~Bissaldi\altaffilmark{12}, 
R.~D.~Blandford\altaffilmark{2}
% ...
}
% ...
\altaffiltext{1}{Deutsches Elektronen Synchrotron DESY, D-15738 Zeuthen, Germany}
\altaffiltext{2}{W. W. Hansen Experimental Physics Laboratory, Kavli Institute for Particle Astrophysics and Cosmology, Department of Physics and SLAC National Accelerator Laboratory, Stanford University, Stanford, CA 94305, USA}
\altaffiltext{3}{Institute for Cosmic-Ray Research, University of Tokyo, 5-1-5 Kashiwanoha, Kashiwa, Chiba, 277-8582, Japan}
\altaffiltext{4}{Universit\`a di Pisa and Istituto Nazionale di Fisica Nucleare, Sezione di Pisa I-56127 Pisa, Italy}
\altaffiltext{5}{Istituto Nazionale di Fisica Nucleare, Sezione di Trieste, I-34127 Trieste, Italy}
\altaffiltext{6}{Dipartimento di Fisica, Universit\`a di Trieste, I-34127 Trieste, Italy}
\altaffiltext{7}{Istituto Nazionale di Fisica Nucleare, Sezione di Padova, I-35131 Padova, Italy}
\altaffiltext{8}{Dipartimento di Fisica e Astronomia ``G. Galilei'', Universit\`a di Padova, I-35131 Padova, Italy}
\altaffiltext{9}{NASA Goddard Space Flight Center, Greenbelt, MD 20771, USA}
\altaffiltext{10}{Department of Physics and Department of Astronomy, University of Maryland, College Park, MD 20742, USA}
\altaffiltext{11}{Istituto Nazionale di Fisica Nucleare, Sezione di Pisa, I-56127 Pisa, Italy}
\altaffiltext{12}{Istituto Nazionale di Fisica Nucleare, Sezione di Bari, I-70126 Bari, Italy}
% ...



\begin{abstract}
On 2015 June 16, \Fermi-LAT observed a giant outburst from the flat spectrum radio quasar 
3C~279 with a peak $>100$\,MeV flux of $\sim3.6\times10^{-5}\;{\rm photons}\;{\rm cm}^{-2}\;{\rm s}^{-1}$ 
averaged over orbital period intervals.
It is the historically highest $\gamma$-ray flux observed from the source including past EGRET observations,
with the $\gamma$-ray isotropic luminosity reaching  $\sim10^{49}\;{\rm erg}\;{\rm s}^{-1}$.
During the outburst, the \Fermi\ spacecraft, which has an orbital period of 95.4 min, was operated in a special pointing mode to optimize the 
exposure for 3C~279.  For the first time, significant flux variability at sub-orbital 
timescales was found in blazar observations by \Fermi-LAT.  The source flux variability was resolved down to 
2-min binned timescales, with flux doubling times less than 5 min.
The observed minute-scale variability suggests a very compact emission region 
at hundreds of Schwarzschild radii from the central engine in conical jet models. 
A minimum bulk jet Lorentz factor 
($\Gamma$) of 35 is necessary to avoid both internal $\gamma$-ray absorption and 
super-Eddington jet power. In the standard external-radiation-Comptonization scenario, 
$\Gamma$ should be at least 50 to avoid overproducing the synchrotron-self-Compton 
component.  However, this predicts extremely low magnetization ($\sim5\times10^{-4}$). 
Equipartition requires $\Gamma$ as high as 120, unless the emitting region is a 
small fraction of the dissipation region.  Alternatively, we consider $\gamma$ rays 
originating as synchrotron radiation of $\gamma_{\rm e}\sim1.6\times10^6$ electrons, 
in magnetic field $B\sim1.3$\,kG, accelerated by strong electric fields $E\sim B$ 
in the process of magnetoluminescence.
At such short distance scales, one cannot immediately exclude production of $\gamma$ rays 
in hadronic processes.  
\end{abstract}
\keywords{galaxies: active --- galaxies: jets --- gamma rays: galaxies ---  
quasars: individual (3C~279) --- radiation mechanisms: non-thermal} 

\section{Introduction}
\label{sec_intro}
Amongst all high-luminosity blazars --- which are active galaxies dominated by 
Doppler-boosted emission from relativistic jets pointing toward our line of 
sight --- 3C~279 is one of the most extensively studied objects.  
This flat spectrum radio quasar (FSRQ: $z = 0.536$) has been 
detected in essentially all accessible spectral bands; 
in particular, strong and variable $\gamma$-ray emission was detected by {\it Compton}/EGRET~\citep{Har92, Kni93}, 
and it was the first FSRQ detected above 100\,GeV~\citep{MAGIC}.
The $\gamma$-ray emission dominates the apparent luminosity of the source, 
and the nature of $\gamma$-ray variability and its relationship to that measured 
in other bands provide the strongest constraints on the total energetics as well 
as the emission processes operating in the jets of luminous blazars~\cite[e.g.,][]{Mar92, Sik94}.

Owing to the all-sky monitoring capability of the {\Fermi} Large Area Telescope~\cite[LAT:][]{LAT}, 
we have a continuous $\gamma$-ray flux history of 3C~279 for more than 7 years.  
3C 279 underwent several outbursts in the past, having flared with a peak $\gamma$-ray 
flux ($E>100$\,MeV) $\sim 10^{-5}\;$\phcms, in 2013 December and 2014 April, 
with fluxes about three times greater than the peak during the first 2 
years of \Fermi-LAT observations~\citep{Hay12, Hay15}. During the flaring 
epoch in 2013 December, the $\gamma$-ray spectrum hardened ($\Gamma_{\gamma}\simeq1.7$ 
in ${\rm d}N/{\rm d}E\propto E^{-\Gamma_{\gamma}}$) and rapid hour-scale flux variability was observed. 
The $\gamma$-ray flux strongly dominated the flux in any other band, indicating a very high 
``Compton dominance'' (the ratio of the total inverse-Compton luminosity over the total synchrotron luminosity) of a factor of 100. 
This in turn suggests extremely low jet magnetization, with a level of $10^{-4}$. 
Those results motivated, e.g., the stochastic acceleration model, 
which could reproduce the hour-scale variability and the hard spectrum 
of the flare event~\citep{FermiII}.

In 2015 June, 3C~279 became very active again with fluxes exceeding the 
2013/2014 level~\citep{Cut15, Pal15b}, and prompting a Target of Opportunity (ToO) 
repointing of \Fermi, resulting in a $\sim2.5$ times greater exposure. 
The measured $\gamma$-ray flux in daily bins reached 
$\sim2.4\times10^{-5}\;$\phcms\ on 2015 June 16, allowing an unprecedented investigation of
variability on timescales even shorter than one {\it Fermi} orbit. 
In this Letter, we report and 
offer an interpretation of the minute-scale variability observed by \Fermi-LAT 
for the first time in {\sl any blazar}.

\section{\Fermi-LAT Gamma-ray Observations}

We analyzed the LAT data following the standard 
procedure\footnote{\texttt{http://fermi.gsfc.nasa.gov/ssc/data/analysis/}}, 
using the LAT analysis software \texttt{ScienceTools} \texttt{v10r01p01} with the 
\texttt{P8R2\_SOURCE\_V6} instrument response functions. 
Events with energies 0.1--300\,GeV were extracted within a $15^{\circ}$ acceptance 
cone Region of Interest (ROI) centered at 3C~279 (RA = $195\fdg047$, 
Dec=$-5\fdg789$, J2000). Gamma-ray spectra were derived by an unbinned 
maximum likelihood fit with \texttt{gtlike}.
The background model included 
sources from the third LAT catalog \citep[3FGL:][]{3FGL} inside the ROI 
and which showed TS$>25$\footnote{``TS'' stands for the test statistic from the likelihood 
ratio test~\citep[see][]{ML}.} based on an analysis of 1-month of LAT data, for 2015 June.
Their spectral parameters were fixed by the fitting results from the 1-month data 
analysis. Additionally, the model included the isotropic and Galactic diffuse 
emission components\footnote{\texttt{iso\_P8R2\_SOURCE\_V6\_v06.txt} and \texttt{gll\_iem\_v06.fits}}\citep{Ace16}, 
with fixed normalizations during the fitting.
Note that the contribution of background components to the 3C~279 flux determinations 
in short-term binned light curves during the outburst is negligible.

\subsection{Light curve}

\begin{figure*}[tbp]
\centering
\includegraphics[height=15cm]{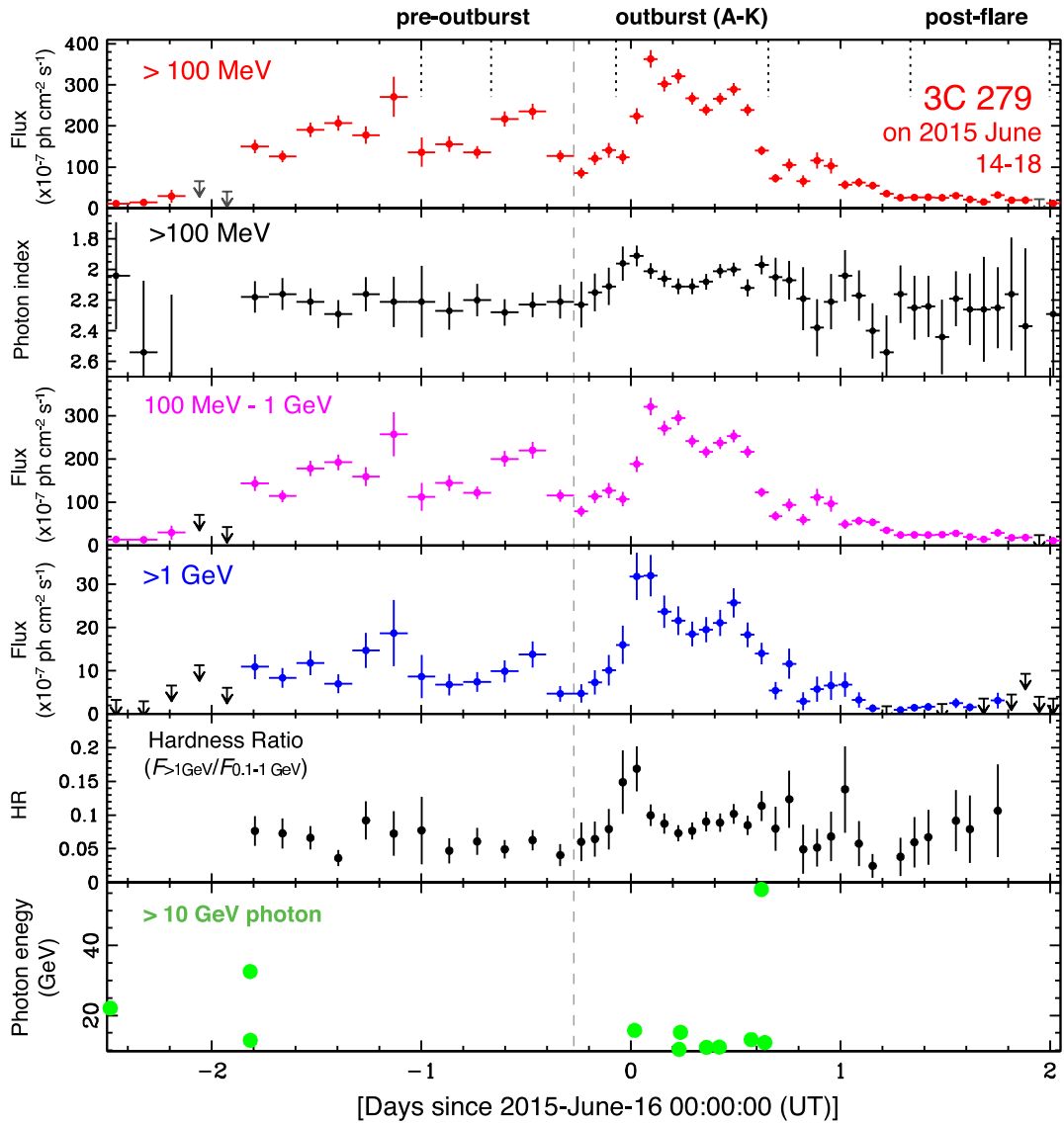}
\caption{Light curves of 3C~279 in the $\gamma$-ray band as observed by \Fermi-LAT.
The vertical dashed line indicates when the \Fermi-LAT observation mode was 
switched from the normal survey mode to the pointing mode of the ToO observations.
The data were binned on a two-orbit timescale (190.8\,min) during the normal survey mode
and on a one-orbit timescale (95.4\,min) during the ToO observations.
The panels from the top to the bottom show: 
(1) integrated flux above 100\,MeV, 
(2) photon index above 100\,MeV, 
(3) integrated flux between 0.1 and 1\,GeV ($F_{0.1-1\;{\rm GeV}}$), 
(4) integrated flux above 1\,GeV ($F_{>1\;{\rm GeV}}$), 
(5) hardness ratio ($F_{>1\;{\rm GeV}}/F_{0.1-1\;{\rm GeV}}$),
(6) arrival time distribution of photons with reconstructed energies above 10\,GeV.  
For bins with TS$<6$, 95\% confidence level upper limits are plotted.
}
\label{LC}
\end{figure*}

Figure~\ref{LC} shows light curves of 3C~279 measured by \Fermi-LAT 
between 2015 June 14 12:00:00 and June 18 00:00:00 UTC, 
including the most intense outburst observed on June 16.
ToO observations were conducted 
from 2015 June 15 17:31:00 through 2015 June 23 16:19:00, 
during which LAT switched from its normal survey mode to 
a pointing mode targeting 3C~279.
For data taken during the normal observing mode, 
the data were binned at a twice the orbital period 
so that individual bins can have more uniform exposure times.
Beginning with the ToO observation, the data were binned orbit by orbit. 
The $\gamma$-ray fluxes and photon indices 
were derived using a simple power-law model.  The hardness ratio in the 
5th panel of Figure~\ref{LC} was defined as the ratio between the hard-band 
($>1$\,GeV) and the soft-band (0.1$-$1\,GeV) fluxes;  
${F_{>1\;{\rm GeV}}/F_{0.1-1\;{\rm GeV}}}$.
Here we define the outburst phase to be between 2015 June 15 22:17:12 and 
June 16 15:46:36 (MJD~57188.92861 and 57189.65736) as 
indicated in Figure~\ref{LC}: it comprises 11 one-orbit 
bins designated Orbit `A' to `K' respectively.

The greatest flux above 100\,MeV was recorded during Orbit~C, centered at 
2015 June 16 02:15:42 (MJD~57189.09424), reaching $(3.6\pm0.2)\times10^{-5}\;$\phcms.
It exceeds the largest 3C~279 flares previously detected by \Fermi-LAT on 2013 December 20, 
2014 April 4~\citep{Hay15, Pal15}, and those detected by EGRET in 1996~\citep{Weh98} 
($\sim1.2\times10^{-5}\;$\phcms), making it the historical largest $\gamma$-ray 
($>100$\,MeV) flare of 3C~279. It is the second-greatest flux among blazars 
observed by \Fermi-LAT after the 3C\,454.3 outburst in 2010 November~\citep{Abd11a}.  
During Orbit~C we found $\Gamma_{\gamma}=2.01\pm0.05$, which was not as hard as $\Gamma_{\gamma}\sim1.7$
observed on 2013 December 20. 
The hardest spectrum during this outburst 
was $\Gamma_{\gamma}=1.91\pm0.07$ in Orbit B.

The highest-energy photon, 56\,GeV, was detected\footnote{probability of association with 3C~279 estimated by \texttt{gtsrcprob} is $>99.99$\%} at 2015 June 16 14:58:12 UTC, 
almost at the end of the outburst phase (Figure~\ref{LC}, bottom), 
corresponding to $\sim15.1$ hours since the outburst began, and $\sim12.7$ hours 
later than the center of the highest-flux time bin.  Interestingly, in the 
2010 November flare of 3C\,454.3, the highest-energy photon (31\,GeV) also arrived 
during the decay part of the main flare~\citep{Abd11a}.

\subsection{Sub-orbital scale variability}

\begin{figure*}[htbp]
\centering
\includegraphics[height=13cm]{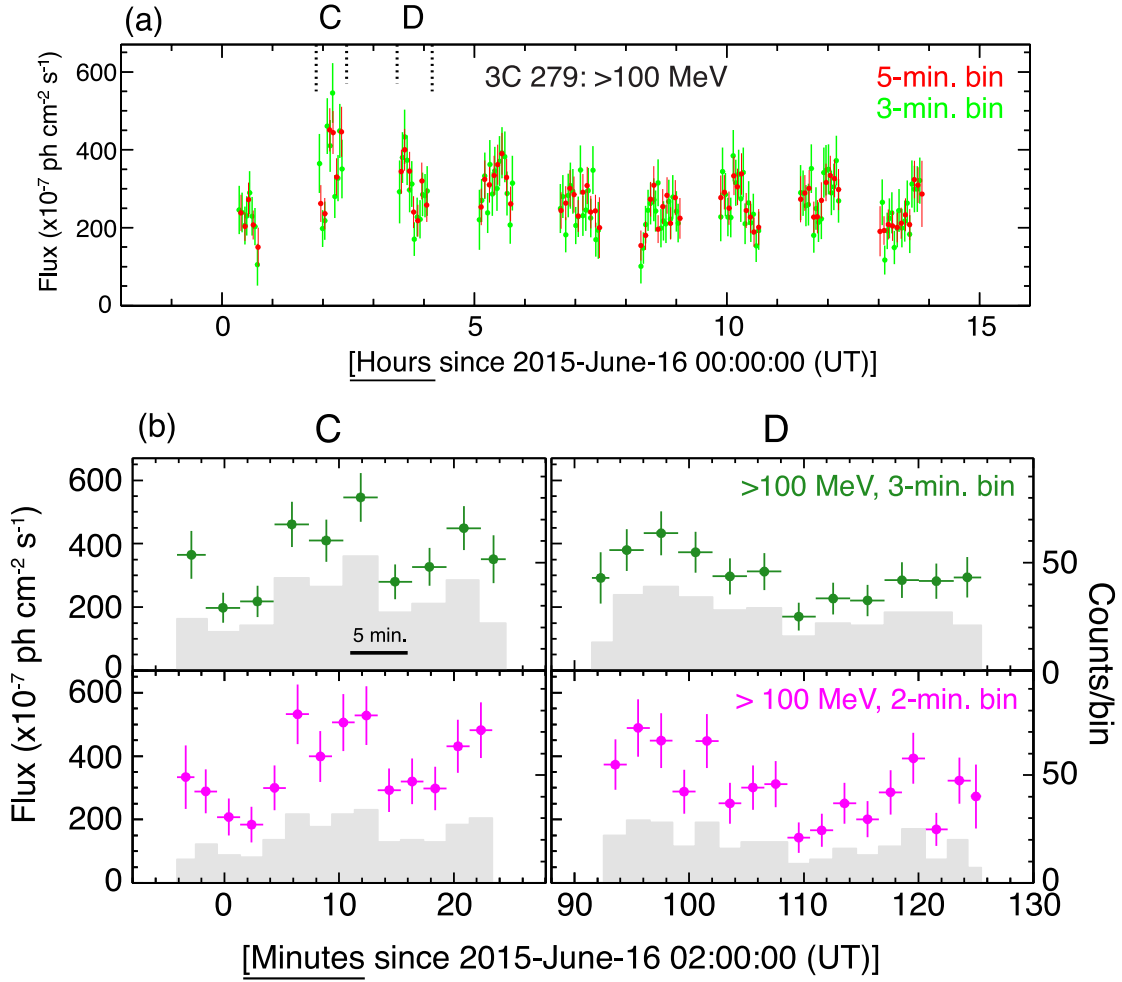}
\caption{Light curves of 3C~279 above 100\,MeV with minute-timescale intervals. (a): Intervals of 5 min (red) and 3 min (green) during the outburst phase from Orbits B--J. 
(b): Enlarged view during Orbits C and D. Each range is indicated with dotted vertical lines in (a). The points denote the fluxes (left axis), and
the gray shaded histograms represent numbers of events (right axis) detected within $8^{\circ}$ radius centered at 3C~279 for each bin. 
Contamination from both diffuse components were estimated as $\sim1$ photon for each 3-min bin.
}
\label{shortLC}
\end{figure*}

The very high $\gamma$-ray flux state and the ToO observations provided a 
sufficiently large number of photons in each bin to resolve light curves 
with shorter timescales than the \Fermi\ orbital period. 
Figure~\ref{shortLC}-(a) shows light curves above 100\,MeV for integration 
times of 5 min (red) and 3 min (green) for Orbits B--J, where the 
orbit-averaged flux exceeded $2\times 10^{-5}\;$\phcms. The spacecraft location and attitude data 
with 1-s resolution were used for analysis of those short-timescale light curves.
To investigate flux variability at sub-orbital periods, we fitted a constant 
value to each orbit for both time bins, 
and calculated a 
probability (p-value) from $\chi^2$ in each orbit.  While many orbits resulted in 
p-values consistent with constant fluxes, we found significant indications 
of variability on a sub-orbital timescale for Orbit C: (p, $\chi^2$/dof)$=$(0.0015, 
19.62/5) and (0.00047, 29.8/9) for 5- and 3-min bins respectively,  
and Orbit D: (p, $\chi^2$/dof)$=$(0.067, 11.79/6) and (0.068, 18.65/11) 
for 5- and 3-min bins, respectively (see details in Table~\ref{tab}).  

Enlarged views of light curves above 100\,MeV for Orbits C and D are in 
Figure~\ref{shortLC}-(b); those show integration times of 3 min and 2 min.
In those time bins, the flux reached $\sim5\times10^{-5}\;$\phcms\ at the highest,
and showed most rapid variations. In the 3-min binned light curve, the flux 
doubled even from the 3rd to the 4th bins, and decreased by almost a half 
from the 6th to the 7th bins. Although defining the characteristic 
timescale of the variability is difficult, the flux doubling time is conservatively less 
than 10~min, and plausibly $\sim5$ min or shorter.


\subsection{Power Density Spectrum}

\begin{figure*}[htbp]
\centering
\plottwo{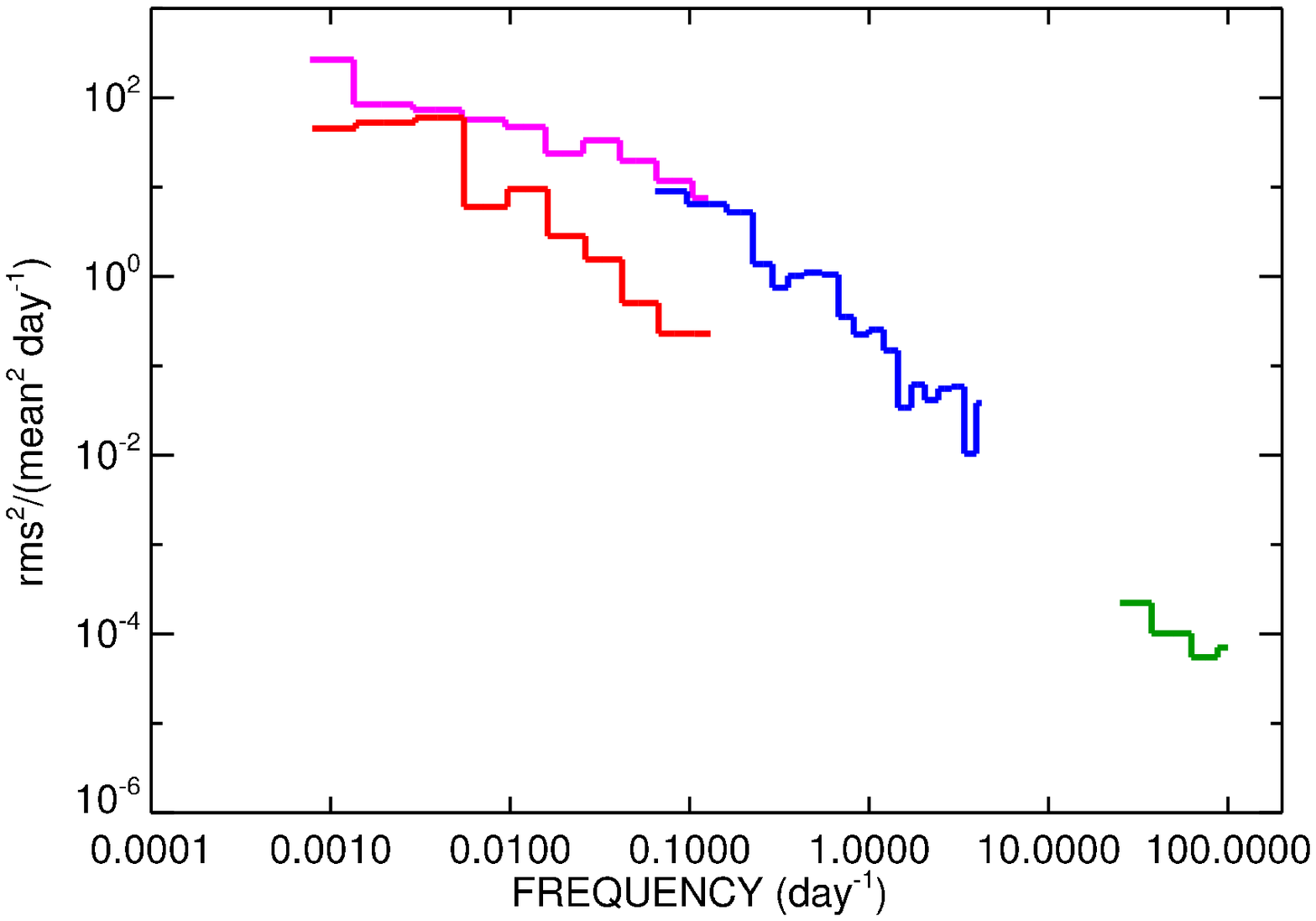}{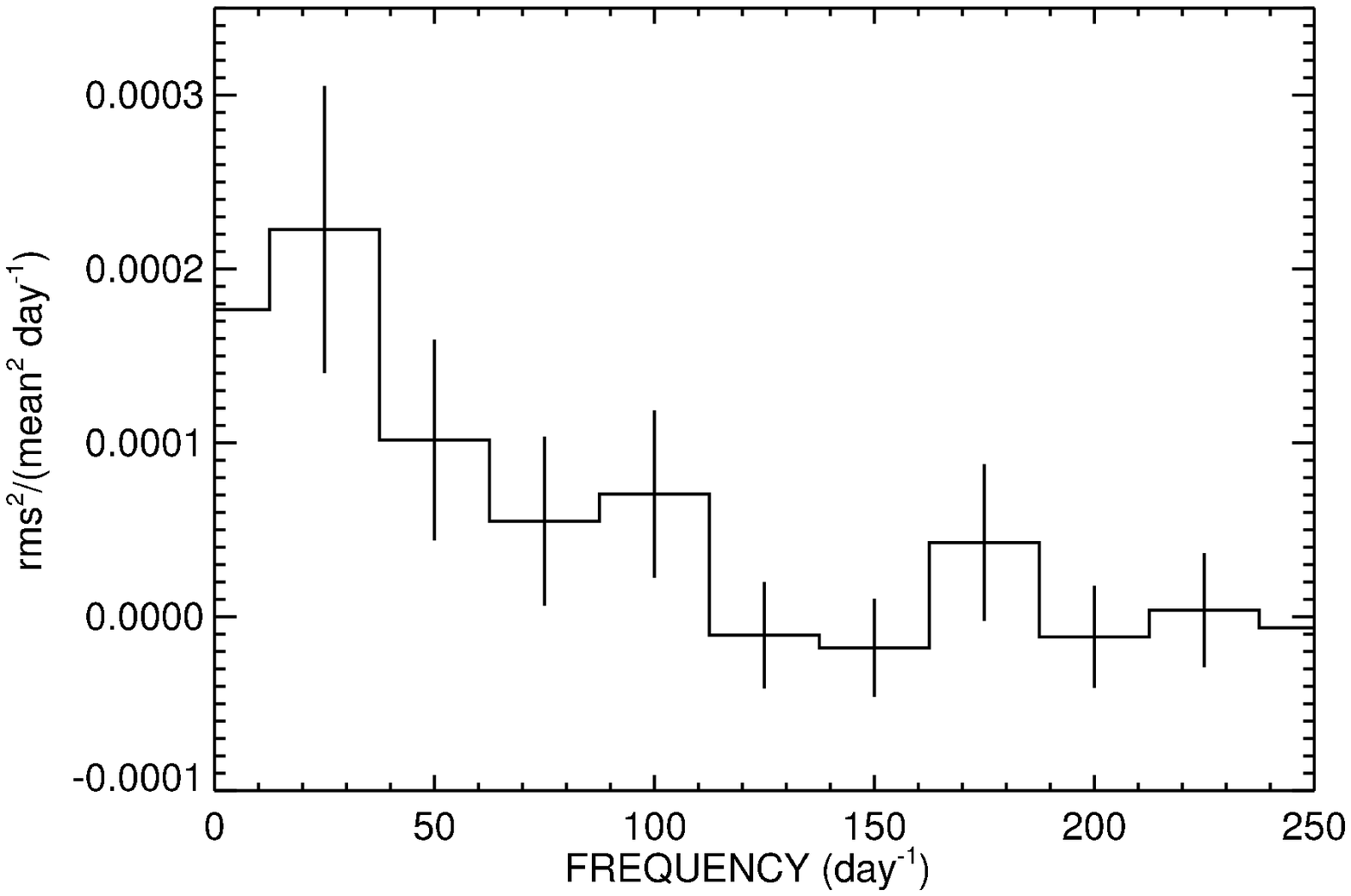}
\caption{ 
Power Density Spectrum (PDS) of the $\gamma$-ray flux of 3C~279.  
(left) PDS derived from three different time-binned light curves:  
3 days (red and magenta), orbital period (blue) and 3 min (green). 
The PDS's marked in red and magenta were derived using the first and second halves of the first 7-year \Fermi-LAT observation, respectively.
The second half of the interval contains the giant outburst phase in 2015 June.
(right) Enlarged view of the high-frequency part of the PDS, 
based on 3-min binned light curves, plotted using a linear scale 
and including also the highest frequencies.
The white noise level has been subtracted.  
}
\label{fig_scal}
\end{figure*}

The available LAT data allow us to study the Power Density Spectrum (PDS) 
on different timescales. Results for three different frequency ranges 
are shown in Figure~\ref{fig_scal}. Two lower-frequency ($<0.1\;{\rm day}^{-1}$) 
PDSs were calculated, each from a 3-day binned light curve covering one half 
of the 7-year LAT data (MJD 54683--55950 and 55950--57254, respectively).  
The PDS for intermediate frequencies is based on a light curve for the 
active period in 2015 June (MJD 57181--57197) binned on the orbital period of \Fermi. 
White noise subtraction was based on the estimated measurement errors in the light 
curves and these were also logarithmically binned before plotting in Figure~\ref{fig_scal}. 
The PDS's for high frequencies were derived from the 3-min binned sub-orbital light 
curves of Orbits B--J.  One PDS was calculated for each orbit, and then these were averaged. 
White noise defined from the flat PDS level above $110\;{\rm days}^{-1}$ has been subtracted. 
The normalization of the PDS means that if the rms/flux is constant during variations 
in source flux, the PDS level will not change. The intermediate frequency PDS connects 
well with the low-frequency PDS for the second 3.5 years, which includes the active period 
in 2015 June.  
The PDS for the second 3.5-year interval shows a higher relative variability 
and a flatter spectrum (slope:$\;-0.61\pm0.06$) compared to the first interval (slope:$\;-1.24\pm0.15$)
as well as a break around $0.1\;{\rm day}^{-1}$.

\subsection{Gamma-ray Spectra}

Gamma-ray spectra measured by \Fermi-LAT, extracted for each orbit during the outburst, 
were fitted to simple power-law (PL) and log-parabola 
(LP: ${\rm d}N/{\rm d}E \propto (E/E_0)^{-\alpha-\beta\log(E/E_0)}$ with $E_0=300\;$MeV) 
models (see Table~\ref{tab}).  The peak energy ($E_{\rm peak}$) of the 
spectral energy distribution (SED) was derived from a fit with the LP model.
Generally, the LP model is more favored than the PL model to describe the spectral shape.
The fitting results suggest that $E_{\rm peak}$ ranges between $\sim300\;$MeV 
and $\sim1\;$GeV during the outburst. At the beginning and end of the outburst,
the spectra appear relatively hard with SED peaks at $\sim1\;$GeV. 
The SED peaks were located at significantly higher energies than for the 
usual states of 3C~279, when the peak is located below the \Fermi-LAT band ($<100\;$MeV),
but lower than $E_{\rm peak}$ observed in the 2013 December 20 flare ($\gtrsim3\;$GeV).

\begin{figure*}[htbp]
\centering
\plotone{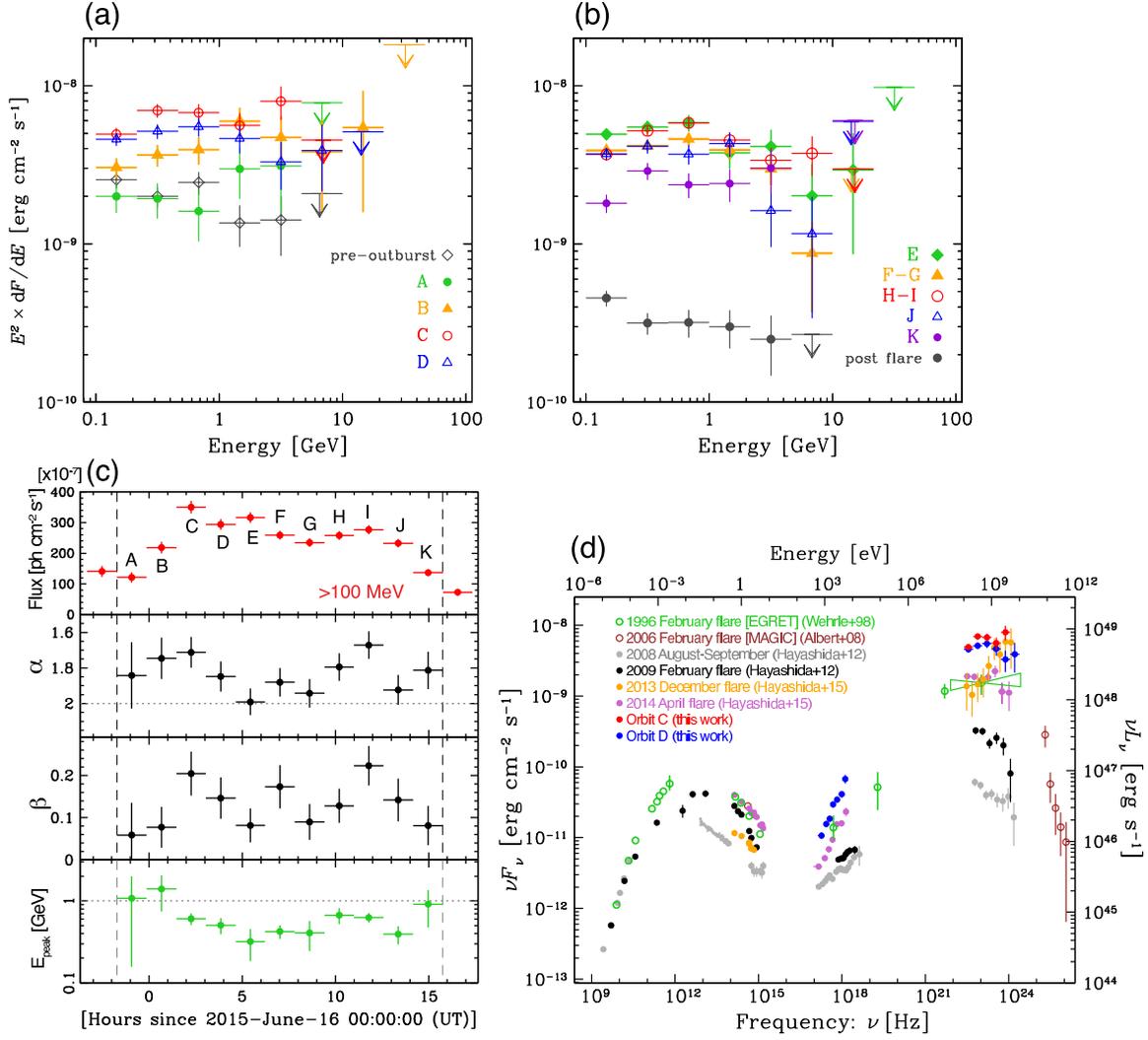}
\caption{(a,b): Gamma-ray SEDs of 3C~279 for each orbit during the outburst phase as well as ``{\it pre-outburst}'' and ``{\it post flare}'' as indicated in Figure~\ref{LC}. The down arrows represent 95\% confidence level upper limits. (c): Best fit parameters of the spectra based on the log-parabola model for each orbit (see Table~\ref{tab} for numbers).
(d) Broad-band SED of Orbit C and D, and some historical multi-band observations with EGRET, MAGIC and \Fermi-LAT.
}
\label{fig:SED}
\end{figure*}

Figure~\ref{fig:SED} shows the $\gamma$-ray SED as measured by \Fermi-LAT for each orbit. 
In these plots, Orbits F and G, and Orbits H and I, were 
combined because they showed similar spectral 
fitting results and fluxes.  The spectra in the ``{\it pre-outburst}'' 
and ``{\it post-flare}'' periods as defined in Figure~\ref{LC} were 
also extracted for comparison. The spectral peaks are apparently 
located within the LAT energy band during the outburst. 
The peak SED flux reaches nearly $\sim10^{-8}\;{\rm erg}\;{\rm cm}^{-2}\;{\rm s}^{-1}$, 
corresponding to an apparent luminosity of $10^{49}\;{\rm erg}\;{\rm s}^{-1}$.

\section{Discussion}

For the first time, \Fermi-LAT detected variability of $>100\;$MeV $\gamma$-ray 
flux from a blazar on timescales of $t_{\rm var,obs} \sim 5\;{\rm min}$ or shorter. 
These timescales are comparable to the shortest variability timescales detected 
above 100\,GeV in a handful of blazars and a radio galaxy by ground-based Cherenkov 
telescopes (PKS~2155$-$304, \citealt{Aha07}; Mrk~501, \citealt{Alb07}; IC~310, \citealt{Ale14}). 
Moreover, this is only the second case when such timescales are reported for an FSRQ 
blazar, after PKS~1222+216~\citep{Ale11}, 
while \Fermi-LAT had only ever detected variability as short as hour timescales 
in some FSRQs~\citep[e.g.,][]{Abd11a, Saito13, Hay15}. 
This observational 
result imposes very stringent constraints on the parameters of the $\gamma$-ray emitting region.

\paragraph{Emitting region size}
The observed variability timescale constrains the characteristic size of the emitting 
region radius $R_\gamma<\mathcal{D}ct_{\rm var,obs}/(1+z)\simeq 10^{-4}(\mathcal{D}/50)\;{\rm pc}$, 
where $\mathcal{D}$ is the Doppler factor.  With such extremely short variability timescale, 
we may consider a significantly larger dissipation region 
(a region from which energy is transferred/supplied to the emitting regions) 
of $R_{\rm diss} = R_\gamma/f_\gamma$, 
where $f_\gamma$ is a scale factor.  Normally, $f_\gamma = 1$; however, values of 
order $f_\gamma\sim0.01-0.1$ are motivated, e.g., by studies of relativistic magnetic 
reconnection \citep{Cer12,Nal12}.  The corresponding characteristic distance scale along 
the jet for a conical geometry is $r_{\rm diss} \simeq R_{\rm diss}/\theta \simeq 
0.005(\Gamma/50)^2(\mathcal{D}/\Gamma)(\Gamma\theta)^{-1}f_\gamma^{-1}\;{\rm pc}$, 
where $\Gamma$ is the Lorentz factor, and $\theta$ is the opening angle.
This corresponds to $\sim100$ Schwarzschild radii ($R_{\rm S}$) for a black hole mass of 
$M_{\rm BH} \sim 5\times 10^8\;{\rm M}_\odot$~\citep[as adopted in][]{Hay15}, and is generally 
well within the broad-line region (BLR) with $r_{\rm BLR} \sim 0.1\;{\rm pc}$ \citep{Tav10}.
While it is typical to assume that $\mathcal{D} \simeq \Gamma$ and $\Gamma\theta \simeq 1$, 
larger distance scales can be obtained for $\Gamma\theta < 1$ \citep[see also][]{Saito15}.  

\paragraph{Jet energetics}
The total jet power required to produce the $\gamma$-ray emission of apparent 
luminosity $L_\gamma \sim 10^{49}\;{\rm erg\,s^{-1}}$ is $L_{\rm j} \simeq L_\gamma/(\eta_{\rm j}\Gamma^2) 
\sim 4\times 10^{46}(\Gamma/50)^{-2}(\eta_{\rm j}/0.1)^{-1}\;{\rm erg\,s^{-1}}$, where $\eta_{\rm j}$ 
is the radiative jet efficiency, typically $\sim0.1$~\citep{Nem12}. The jet power will exceed the Eddington luminosity of 
$L_{\rm Edd} \sim 8\times 10^{46}\;{\rm erg\,s^{-1}}$ for $\Gamma < 35(\eta_{\rm j}/0.1)^{-1/2}$.

\paragraph{Internal absorption}
The optical depth for internal $\gamma$-ray absorption is given by 
$\tau_{\rm \gamma\gamma,int}\simeq(1+z)^2\sigma_{\rm T}L_{\rm soft}E_{\rm max,obs}/(72
\pi m_{\rm e}^2c^6\mathcal{D}^6t_{\rm var})$~\citep{Don95,Beg08}. 
Excluding a single $\simeq 56\;{\rm GeV}$ photon, several photons were detected 
during the outburst with energies in the range 10--$15\;{\rm GeV}$, and hence we 
adopt a maximum photon energy of $E_{\rm max,obs}=15\;{\rm GeV}$.
Based on the \Swift-XRT observation performed during Orbit D (obsID 35019180), 
which resulted in a high source flux of $(5.5\pm0.2)\times10^{-11}\,{\rm erg}\,{\rm 
cm}^{-2}\,{\rm s}^{-1}$ with a hard photon index of $\Gamma_{\rm X}=1.17\pm0.06$ (for the 
0.5--$5\;$keV band) and $L_{\rm X}\sim10^{47}\,{\rm erg}\,{\rm s}^{-1}$,  
a soft radiation ($\sim17\;$keV) luminosity of $L_{\rm soft} \sim 3\times 
10^{47}\;{\rm erg\,s^{-1}}$ has been adopted.
The minimum Doppler factor corresponding to $\tau_{\rm \gamma\gamma,int} = 1$ 
is $\mathcal{D}_{\rm min,int}\simeq25$.

\paragraph{External absorption}
External radiation fields can absorb the $\gamma$-ray photons observed at 
$E_{\rm max,obs}$ when $E_{\rm ext}>2(m_{\rm e}c^2)^2/[(1-\mu)(1+z)E_{\rm max,obs}] 
\simeq 23/(1-\mu)\;{\rm eV}$ in the source frame 
($\mu=\cos{\theta_{\rm scat}},\;{\theta_{\rm scat}}$:scattering angle). 
At short distance scales
$R_{\rm diss} \ll r_{\rm BLR}$, additional absorption may arise from the UV 
or soft X-ray radiation produced by the accretion disk or its corona 
\citep{Der92}, radiation reprocessed by the surrounding medium \citep{Bla95}, 
or from high-ionization He\,{\sc ii} lines \citep{Pou10}. 
However, the observed photon statistics are insufficient to derive quantitative results for the absorption.

\paragraph{ERC scenario}
In the standard leptonic model of FSRQs, $\gamma$ rays are produced by the 
External Radiation Comptonization (ERC) mechanism \citep[e.g.,][]{Sik09}.
This requires that leptons are accelerated to Lorentz factors of $\gamma_{\rm e} 
\simeq [(1+z)E_{\rm obs}/(\mathcal{D}\Gamma E_{\rm ext})]^{1/2} 
\simeq 250\,(\Gamma/50)^{-1}(E_{\rm ext}/10\;{\rm eV})^{-1/2}$.
The radiative cooling timescale satisfies $t_{\rm cool}'(\gamma_{\rm e}) \simeq 
3m_{\rm e}c/(4\sigma_{\rm T}\gamma_{\rm e}u_{\rm ext}')\lesssim t_{\rm var}'$ for the 
minimum energy density of external radiation fields $u_{\rm ext,min}' 
\simeq 3m_{\rm e}c/(4\sigma_{\rm T}t_{\rm var,obs})[(1+z)E_{\rm ext}/E_{\rm cool,obs}]^{1/2} 
\sim 40\;{\rm erg\,cm^{-3}}$. This minimum energy density can be provided by broad emission lines 
\citep[$u_{\rm BLR}'\simeq0.37\Gamma^2 \xi_{\rm BLR}\;$erg\,cm$^{-3}$;][]{Hay12}
for $\Gamma>33$, assuming a covering factor of $\xi_{\rm BLR}\sim0.1$. 
The same leptons would also produce a synchrotron component peaking in the 
mid-IR band at luminosity $L_{\rm syn}\sim L_\gamma/q$, where $q\sim100$ is the Compton dominance,
and a synchrotron self-Compton (SSC) component peaking in the hard X-ray band at luminosity 
$L_{\rm SSC}\sim L_\gamma^2/(4\pi cq\mathcal{D}^4R_\gamma^2u_{\rm ext}')$~\citep{Nal14}.
$L_{\rm SSC}$ would exceed the X-ray luminosity observed by \Swift-XRT 
for jet Lorentz factors $\Gamma<46$. The magnetic jet power can 
be estimated as $L_{\rm B}=\pi R_{\rm diss}^2\Gamma^2u_{\rm BLR}'c/q\simeq 
2\times 10^{43}(\Gamma/50)^6f_\gamma^{-2}\;{\rm erg\,s^{-1}}\sim0.0005(\Gamma/50)^8f_\gamma^{-2}(\eta_{\rm j}/0.1)L_{\rm j}$.
Hence, the jet Lorentz factor $\Gamma=50$, while satisfying the Eddington, 
opacity, cooling, and SSC constraints --- and already much higher than the 
value inferred from radio observations, $\Gamma_{\rm var}
\simeq21$~\citep{Hov09} --- corresponds to severe matter domination.
However, since $L_{\rm B}/L_{\rm j}\propto\Gamma^8$, equipartition defined 
as $L_{\rm B}=L_{\rm j}/2$~\citep{Der14} can be obtained for $\Gamma_{\rm eqp}\simeq120f_\gamma^{1/4}$.

\paragraph{Synchrotron scenario}
Alternatively, we consider a more exotic scenario, in which $\gamma$ rays are 
produced as synchrotron radiation by energetic electrons in a strong magnetic field 
--- in addition to the standard synchrotron component produced under typical 
conditions --- motivated by the $\gamma$-ray flares of the Crab Nebula \citep{Abd11b}, 
and also investigated in the context of the $100\;{\rm GeV}$ emission from FSRQ 
PKS~1222+216 \citep{Nal12}. This scenario requires leptons accelerated to 
Lorentz factors $\gamma_{\rm e}\simeq [(1+z)E_{\rm obs}/(20{\rm neV}\times\mathcal{D}B')]^{1/2} 
\simeq1.6\times 10^6\,(E_{\rm obs}/1\,{\rm GeV})^{1/2}(\Gamma/25)f_\gamma^{-1/2}$, 
where the magnetic field strength can be estimated from the equipartition magnetic 
jet power as $B'\simeq(1+z)f_\gamma(8L_{\rm B}/c)^{1/2}/(\Gamma^2ct_{\rm var,obs}) 
\simeq1.3\;{\rm kG}\times(\Gamma/25)^{-3}f_\gamma$. With such energetic leptons, 
the inverse-Compton scattering proceeds in the Klein-Nishina regime, and 
neither SSC or ERC components are important. A high bulk Lorentz factor $\Gamma\simeq25$ 
is still required for avoiding internal absorption of $\gamma$ rays
and it is helpful for pushing the observed synchrotron photon energy limit to 
comfortable values $E_{\rm syn,max}\simeq4\;{\rm GeV}\times(\Gamma/25)(E'_\parallel/B'_\perp)$ 
\citep[e.g.,][]{Cer12}.  Particle acceleration in magnetic reconnection sites with 
$E'_\parallel > B'_\perp$ \citep{Kir04,Uzd11} is not necessary. The synchrotron cooling 
timescale is $\sim3\;{\rm ms}$ in the co-moving frame, placing this scenario in 
the fast cooling regime.  This will result in a low-energy electron tail 
$N(\gamma_{\rm e})\propto\gamma_{\rm e}^{-2}$, and a corresponding spectral tail $EF(E)\propto E^{0.5}$.

\paragraph{Hadronic scenarios}
On the very small distance scales, the radiative efficiency of the proton-initiated 
cascade mechanism \citep{Man92} is enhanced due to very dense target radiation 
fields, and that of the proton-synchrotron mechanism \citep{Aha00} is enhanced 
due to very strong magnetic fields ($B'\gtrsim{\rm kG}$). A careful analysis of 
these mechanisms, including the non-linear feedback effects \citep{Pet12}, requires a dedicated study.

\paragraph{Dissipation mechanism}
The observed variability timescale and luminosity require extremely efficient 
bulk jet acceleration within $\sim 100\;R_{\rm S}$.  In the synchrotron scenario, 
they also require extremely efficient particle acceleration, going beyond the 
established picture of the blazar sequence \citep{Ghi98}. Magnetic reconnection 
was invoked as a dissipation focusing mechanism effectively increasing the scale 
of the dissipation region by $f_\gamma^{-1}\sim10$--100 \citep{Cer12}.  Relativistic 
reconnection can also produce relativistic outflows dubbed `minijets', which can 
provide additional local Lorentz boost \citep{Gia09}.  In general, magnetic dissipation can 
lead to rapid conversion of magnetic energy to radiation by the process called 
{\sl magnetoluminescence} \citep{Bla15}.

\section{Summary}
In this Letter, we reported the first minute-timescale $\gamma$-ray flux 
variability observed by \Fermi-LAT in an FSRQ blazar, 3C\,279.  In the standard 
ERC scenario with conical jet geometry, the minute-scale variability requires 
a high $\Gamma$ ($>50$) and extremely low magnetization even at the jet base 
($\sim100\,R_S$) or $\Gamma\sim120$ under equipartition. 
The high $\Gamma$ and/or low magnetization at the jet base pose challenges to
standard models of electromagnetically driven jets. 
We also discuss an alternative, synchrotron origin for the GeV 
$\gamma$-ray outburst, which would work in a magnetically dominated jet, 
but requires higher 
electron energies and still implies $\Gamma\sim25$ at the jet base.  

\acknowledgements
The \textit{Fermi}-LAT Collaboration acknowledges support for LAT development, 
operation and data analysis from NASA and DOE (United States), CEA/Irfu and 
IN2P3/CNRS (France), ASI and INFN (Italy), MEXT, KEK, and JAXA (Japan), and 
the K.A.~Wallenberg Foundation, the Swedish Research Council and the National 
Space Board (Sweden). Science analysis support in the operations phase from 
INAF (Italy) and CNES (France) is also gratefully acknowledged. 


\begin{deluxetable*}{c|cccccccccc}
\tablecaption{Flux and spectral fitting results of 3C~279 above 100\,MeV for each orbit (A$-$K) during the outburst phase.\label{tab}}.
\centering                         
\startdata       
\hline \hline
Orbit & Flux\tablenotemark{a} & $\Gamma_{\gamma}$ & $\alpha$ & $\beta$ & $E_{\rm peak}$ & TS & $-2\Delta L$\tablenotemark{c} & p-value\tablenotemark{d}  & p-value\tablenotemark{d}  & $E_{\rm max}$ \\
number & ($10^{-7}$)  & (PL\tablenotemark{b}) & (LP\tablenotemark{b}) & (LP\tablenotemark{b}) & (GeV) &  & &  (5 min.~bin) & (3 min.~bin) & (GeV) \\
\hline 
A &	$121\pm17$ &	$1.96\pm0.11$ & $1.84\pm0.19$ &	$0.06\pm0.08	$ & $1.1\pm0.9$  & 502 & 0.7 & \nodata & \nodata & 8.8 \\
B &	$218\pm19$	& $1.91\pm0.07$	 & $1.75\pm0.12$	& $0.08\pm0.05$	& $1.40\pm0.66$	 & 1346 &	3.2 &	0.434 &	0.453 & 16.9 \\
C &	$350\pm21$	& $2.01\pm0.05$	&  $1.71\pm0.09$	& $0.20\pm0.05$	& $0.61\pm0.10$	& 3037 & 21.9 &0.00148 &0.000474 & 9.3 \\
D &	$294\pm18$	& $2.06\pm0.05$	 & $1.85\pm0.09$	& $0.15\pm0.05$	& $0.50\pm0.11$	 & 2661	& 11.8 &	0.0668	 &0.0677 & 6.7 \\
E &	$316\pm17$	& $2.11\pm0.05$	 & $1.99\pm0.08$	& $0.08\pm0.04$	& $0.32\pm0.14$	 & 3400	& 5.0 &	0.504	 & 0.429 & 15.2\\
F &	$259\pm14$	& $2.11\pm0.05$	 & $1.88\pm0.08$	& $0.17\pm0.06$ & 	$0.42\pm0.08$	 & 3036	& 15.6	& 0.902	 & 0.419 & 9.2 \\ 
G &	$235\pm14$	& $2.08\pm0.05$	 & $1.94\pm0.08$	& $0.09\pm0.04$ &	$0.41\pm0.16$	 & 2720	& 5.6 & 	0.166	 & 0.308 & 10.9 \\
H &	$258\pm14$	& $2.01\pm0.05$	  & $1.79\pm0.08$	& $0.13\pm0.04$	& $0.67\pm0.15$	 & 3309	& 13.4	& 0.228	 & 0.140 & 10.9 \\
I &	$277\pm15$	& $2.00\pm0.04$	 & $1.67\pm0.08$	& $0.22\pm0.05$	& $0.63\pm0.08$	& 3699	& 32.8 &	0.708	 & 0.435 & 7.7 \\
J &	$233\pm14$	& $2.12\pm0.05$	 & $1.92\pm0.08$	& $0.14\pm0.05$	& $0.39\pm0.10$	& 2630	& 10.3 &	0.404	& 0.177 & 13.1\\
K &	$137\pm11$	& $1.97\pm0.06$	 & $1.81\pm0.11$	& $0.08\pm0.05$	& $0.91\pm0.44$	 & 1540	& 3.8	& \nodata & \nodata & 56.0 
\enddata
\tablenotetext{a}{Orbit-averaged flux above 100\,MeV in {\phcms}.}
\tablenotetext{b}{PL: power-law model, LP: log-parabola model.}
\tablenotetext{c}{$\Delta L$ represents the difference of the logarithm of the total likelihood of the fits between PL and LP models.}
\tablenotetext{d}{p-value based on $\chi^2$ fits with a constant flux to each orbit for 5-min.\ and 3-min.\ binned light curves.}
\end{deluxetable*}

\end{document}